\documentstyle[aps,epsf]{revtex}

\begin{document}

\title {Nondecay probability of a metastable state:
       almost exact analytical description \\
       in wide range of noise intensity}

\author{ Svetlana P. Nikitenkova \\
\small{Nizhny Novgorod State Technical University, Applied
Mathematics Dpt.,\\
24 Minin str., Nizhny Novgorod, 603155, Russia.
E-mail: spn@waise.nntu.sci-nnov.ru \\}}
\author{Andrey L. Pankratov  \\
\small{Institute for Physics of Microstructures
of RAS, GSP 105, \\ Nizhny Novgorod, 603600,
RUSSIA. E-mail: alp@ipm.sci-nnov.ru\\}}

\maketitle

\begin{abstract}
This paper presents a complete
description of noise-induced decay of a metastable state in a
wide range of noise intensity. Recurrent formulas of exact
moments of decay time valid for arbitrary noise intensity have
been obtained.
The nondecay probability of a metastable state was
found to be really close to the exponent even for the case when
the potential barrier height is comparable or smaller
than the noise intensity.
\end{abstract}
\section{Introduction}
Overdamped Brownian motion in a field of force (Markov process)
is a model widely used
for description of noise-induced transitions in different
polystable systems. In many practical tasks (e.g. tasks of
Josephson electronics, kinetics of chemical reactions and so on)
it is enough to know the probability of transition or even only
a time scale of transition. When the transition occurs over a
potential barrier high enough in comparison with noise
intensity, the probability of transition is a simple exponent
$\sim\exp(-t/\tau)$ \cite{G}, where $\tau$ is the mean transition time.
In this case the mean transition time gives complete information
about the probability evolution. The boundaries of validity of exponential
approximation of the probability were previously studied in
\cite{Sch},\cite{Nad}. In \cite{Sch} authors extended the Mean
First Passage Time to the case of "radiation" boundary condition
and for two barrierless examples demonstrated good coincidence
between exponential approximation and numerically obtained
probability. In a more general case the exponential behavior of observables
was demonstrated in \cite{Nad} for relaxation processes in
systems having steady states. Using the approach of
"generalized moment approximation" the authors of \cite{Nad}
obtained the exact mean relaxation time to steady state and for
particular example of a rectangular potential well demonstrated
good coincidence of exponential approximation with numerically
obtained observables.
The considered in \cite{Nad} example of the rectangular well
does not have a potential barrier, and the authors of that paper supposed
that their approach (and the corresponding formulas) should also give
good approximation in tasks with diffusive barrier crossing for
a wide range of noise intensity.

In the frame of this paper we consider a different case than in
\cite{Sch},\cite{Nad}: we present investigation of nondecay
probability of a metastable state. We treat the decay as a transition
of Markov process trajectory outside the region of a metastable state.
We consider namely the probability to find a relization of Markov
process in a given interval, but not the probability to pass
the boundary for the first time (First Passage Time formalizm),
since usually in experiment we can measure only the first one.
In most of applied tasks no absorbing boundaries can be
introduced and the precision of the used equipment allows to
register the only fact of leaving the considered domain,
but the boundary may be crossed many times during the transition
(infinite number of times for Markov process, since this process
is not differentiable).

Using the approach proposed by Malakhov \cite{MP},\cite{M},
that requires only knowledge of the behavior of a potential at
$\pm\infty$, we have decomposed the nondecay probability into a
set of moments (cumulants), obtained recurrent formulas for these
moments and approximately summarized them into the required
probability. The obtained nondecay probability demonstrates
exponential behavior with a good precision even in the case of a
small potential barrier in comparison with noise intensity.

\section{Main equations and set up of the problem}

    Consider a process of Brownian diffusion in a potential profile
${\mit\Phi(x)}$. Let a coordinate $x(t)$ of the Brownian particle
described by the probability density $W(x,t)$ at initial instant
of time has a fixed value $x(0)=x_0$ within the interval $(c,d)$,
i.e. the initial probability density is the delta function:
$W(x,0)= \delta(x-x_0)$, $x_0\in (c,d)$.

      In this case the one-dimensional probability
density $W(x,t)$ is the transition probability density from the point
$x_0$ to the point $x$: $W(x,t)=W(x,t;x_0,0)$.
It is known that the probability density $W(x,t)$ of the Brownian
particle in the overdamped limit satisfies to the Fokker--Planck equation
(FPE):
\begin{eqnarray}
{\partial W(x,t)\over\partial t}=
-{\partial G(x,t)\over\partial x}=
{1\over B}\left\{{\partial\over\partial x}\left[
{d\varphi(x)\over dx}W(x,t)\right]+
{\partial^2W(x,t)\over\partial x^2}\right\}. & &\label{2}
\end{eqnarray}
with the delta-shaped initial distribution.
Here $B=\displaystyle{h/{kT}}$, $G(x,t)$ is the probability
current, $h$ is the viscosity (in computer simulations we put
$h=1$), $T$ is the temperature, $k$ is  the Boltzmann constant and
$\varphi(x)=\displaystyle{\mit\Phi(x)/kT}$ is the dimensionless potential
profile. In this paper we restrict ourselves by the case of metastable
potentials, i.e. we consider an overdamped Brownian motion in a potential
field $\varphi(x)$ in systems, having metastable states, such that
${\varphi(-\infty)}=+\infty$ and ${\varphi(+\infty)}=-\infty$.
This leads to the following boundary conditions:
$G(-\infty,t)=W(+\infty,t)=0$.
Note, that the results obtained may be generalized for potentials
of arbitrary types, e.g. for such that $\varphi(\pm\infty)=\infty$.

     It is necessary to find the probability $P(x_0,t)$
of a Brownian particle, located at the point $x_0$ ($t=0$) within
the interval ($c,d$) to be at the time $t>0$ inside the
considered interval:
$P(x_0,t)=\int\limits_{c}^d W(x,t)dx$. Further we for simplicity
will call the probability $P(x_0,t)$ as nondecay probability.
We suppose, that $c$ and $d$ are arbitrary chosen points of an arbitrary
potential profile ${\varphi(x)}$ and boundary conditions at these points
may be arbitrary: $W(c,t)\ge 0$, $W(d,t)\ge 0$. In this case there is the
possibility for a Brownian particle to come back in the interval ($c,d$)
after crossing boundary points.

\section{Moments of decay time}

     Consider the nondecay probability $P(x_0,t)$. We can
decompose this probability to the set of moments. On the other
hand, if we know all moments, we can in some cases construct a
probability as the set of moments. Thus, analogically to moments
of the First Passage Time \cite{Pon}-\cite{ris} we can introduce
moments of decay time $\tau_n(c,x_0,d)$ (or, generally, moments of
transition time, see \cite{P}, where it was
performed for the probability $Q(x_0,t)=1-P(x_0,t)$):
\begin{eqnarray}\label{3}
\tau_n(c,x_0,d)=<t^n>={\int\limits_0^{\infty} t^n
{\partial P(x_0,t)\over\partial t}dt\over{P(x_0,\infty)-P(x_0,0)}}.
\end{eqnarray}

    Here we can formally denote the derivative of the probability
divided by the normalization factor as $w(x_0,t)$ and thus
introduce the probability density of decay time $w(x_0,t)$
in the following way \cite{P}:
\begin{eqnarray}\label{2a}
w(x_0,t)={\partial P(x_0,t)\over{\partial{t}}} \frac{1}{[P(x_0,\infty)-
P(x_0,0)]}.
\end{eqnarray}
It is important to mention that the moments of
decay (transition) time (\ref{3}) is a
generalization of the well-known moments of the First Passage
Time for the case of arbitrary boundary conditions (see
discussion in \cite{P}). For example, in the considered
case of the potential $\varphi(x)$ (such that ${\varphi(-\infty)}=+\infty$
and ${\varphi(+\infty)}=-\infty$) the moments of decay time coincide
with the corresponding moments of the First Passage Time, if a
reflecting boundary at the point $c$ and an absorbing boundary at
the point $d$ are introduced. On the other hand, if we consider
the decay of metastable state as transition over a barrier top,
and compare mean decay time obtained via approach discussed in the
present paper (case of a smooth potential without absorbing boundary)
and the mean First Passage Time (MFPT) of the absorbing boundary
located at the barrier top, we get two times difference between
these time characteristics even in the case of a high potential
barrier in comparison with the noise intensity.
This is due to the fact, that the MFPT does not take into account
the backward probability current and therefore is sensitive
to the location of an absorbing boundary.
For the considered situation, if we will move the
boundary point down from the barrier top, the MFPT will increase
up to two times and tend to reach value of the corresponding mean
decay time, which is less sensitive to the location of the
boundary point over a barrier top. Such weak dependence of the
mean decay time from the location of the boundary point at the
barrier top or further is intuitively obvious: much more time
should be spent to reach the barrier top (activated escape) than
to move down from the barrier top (dynamic motion).

The required moments of decay time may be obtained via the approach
proposed by Malakhov \cite{MP},\cite{M}. This approach is based
on the Laplace transformation method of the FPE (\ref{2}).
Following this approach, one can
introduce the function $H(x,s)\equiv s\hat{G}(x,s)$,
where $\hat{G}(x,s)=\int\limits_0^\infty G(x,t) e^{-st}dt$
is the Laplace transformation of the probability current, and
expand it in the power series in $s$:
\begin{equation}
H(x,s)\equiv s\hat{G}(x,s)=H_0(x)+sH_1(x)+s^2H_2(x)+ \ldots
\label{a1}
\end{equation}

It is possible to find the differential equations for $H_n(x)$
(see \cite{MP},\cite{M}; $dH_0(x)/dx=0$):
\begin{equation}
\begin{array}{ll}
\displaystyle{\frac{dH_1(x)}{{dx}}}=\delta(x-x_0), &  \\
\displaystyle{\frac{d^2 H_n(x)}{{dx^2}}}+{\frac{d\varphi(x)}{{
dx}}}{\frac{dH_n(x)}{{dx}}}=BH_{n-1}(x),
\quad n=2,3,4,\ldots  \label{a3}
\end{array}
\end{equation}
Using the boundary conditions $W(+\infty,t)=0$ and $G(-\infty,t)=0$,
one can obtain from (\ref{a3}) $H_1(x)=1(x-x_0)$ and

\begin{eqnarray}
\begin{array}{ll}
H_2(x)=-B\int\limits_{-\infty}^{x}e^{-\varphi(v)}
\int\limits_{v}^{\infty}e^{\varphi(y)}1(y-x_0) dy dv, & \\
H_n(x)=-B\int\limits_{-\infty}^{x}e^{-\varphi(v)}
\int\limits_{v}^{\infty}e^{\varphi(y)}H_{n-1}(y) dy dv,
\quad n=3,4,5,\ldots
\end{array}
\label{rec}
\end{eqnarray}

Why did we calculate this recurrent formula
for the functions $H_n(x)$?
The matter is, that from formula (\ref{3}) (taking the integral
by parts and Laplace transforming it using the property
$P(x_0,0)-s\hat{P}(x_0,s)=\hat{G}(d,s)-\hat{G}(c,s)$ together
with the expansion (\ref{a1})) one can get the following expressions
for moments of decay time:
\begin{equation}
\begin{array}{llll}
\tau_1(c,x_0,d)=-(H_2(d)-H_2(c)), \\
\tau_2(c,x_0,d)=2(H_3(d)-H_3(c)), \\
\tau_3(c,x_0,d)=-2\cdot 3(H_4(d)-H_4(c)),\dots \\
\tau_n(c,x_0,d)=(-1)^n n! (H_{n+1}(d)-H_{n+1}(c)).
\end{array}
\label{t}
\end{equation}

One can represent the $n$-th moment in the following form:

\begin{equation}\label{9}
\tau_n(c,x_0,d)=n!\tau_{1}^{n}(c,x_0,d)+r_n(c,x_0,d).
\end{equation}
This is a natural representation of $\tau_n(c,x_0,d)$ due to
the structure of recurrent formulas (\ref{rec}), which is seen from
the particular form of the first and the second moments for the
case $c=-\infty$ ($c<x_0<d$). From the recurrent formulas
(\ref{rec}), (\ref{t}) one can obtain:
\begin{equation}\label{t1}
\tau_1(-\infty,x_0,d)=B\left\{\int\limits_{-\infty}^d
e^{-\varphi(x)}dx\cdot\int\limits_{x_0}^{\infty}
e^{\varphi(v)}dv-\int\limits_{x_0}^{d}
e^{-\varphi(x)}\int\limits_{x_0}^x e^{\varphi(v)}dvdx\right\}.
\end{equation}
\begin{eqnarray}
\tau_2(-\infty,x_0,d)=2B^2\left\{\left[\tau_1(-\infty,x_0,d)
\right]^2\right.+\nonumber \\
+\int\limits_{-\infty}^d e^{-\varphi(x)}dx\cdot
\int\limits_{x_0}^{\infty} e^{\varphi(v)}
\int\limits_{d}^{v} e^{-\varphi(u)}
\int\limits_{u}^{\infty} e^{\varphi(z)}dzdudv- \label{t2}\\
-\left.\int\limits_{x_0}^{d} e^{-\varphi(x)}
\int\limits_{x_0}^x e^{\varphi(v)}
\int\limits_{d}^{v} e^{-\varphi(u)}
\int\limits_{u}^{\infty} e^{\varphi(z)}dzdudvdx\right\}.
\nonumber
\end{eqnarray}

Using the approach, applied in the paper by Shenoy and Agarwal
\cite{SA} for analysis of moments of the First Passage Time, it can be
demonstrated, that in the limit of a high barrier
$\Delta\varphi\gg 1$ ($\Delta\varphi=\Delta{\mit\Phi}/kT$ is the
dimensionless barrier height) the remainders $r_n(c,x_0,d)$
in formula (\ref{9}) may be neglected. For
$\Delta\varphi\approx 1$, however, a rigorous analysis should
be performed for estimation of $r_n(c,x_0,d)$. Let us suppose,
that the remainders $r_n(c,x_0,d)$ may be neglected in wide range
of parameters and further we will check numerically
when our assumption is valid.

The cumulants of decay time $\ae_n$ \cite{cum},\cite{ris} are
much more useful for our purpose to construct the probability
$P(x_0,t)$, that is the integral transformation of the
introduced probability density of decay time $w(x_0,t)$ (\ref{2a}).
Unlike the representation via moments, the Fourier transformation
of the probability density (\ref{2a}) - the characteristic
function - decomposed into the set of cumulants may be inversely
transformed into the probability density.

Analogically to representation for moments (\ref{9}), similar
representation can be obtained for cumulants $\ae_n$:
\begin{equation}\label{10}
\ae_n(c,x_0,d)=(n-1)!\ae_{1}^{n}(c,x_0,d)+R_n(c,x_0,d).
\end{equation}
It is known that the characteristic function $\Theta(x_0,\omega)
=\int\limits_0^\infty w(x_0,t)e^{j\omega t}dt$ ($j=\sqrt{-1}$)
can be represented as the set of cumulants ($w(x_0,t)=0$ for $t<0$):
\begin{equation}\label{char}
\Theta(x_0,\omega)=\exp\left[\sum_{n=1}^{\infty}
{\ae_n(c,x_0,d)\over n!}(j\omega)^n\right].
\end{equation}

In the case, when the remainders $R_n(c,x_0,d)$ in (\ref{10})
(or $r_n(c,x_0,d)$ in (\ref{9}))
may be neglected, the set (\ref{char})
may be summarized and inverse Fourier transformed:
\begin{equation}\label{11}
w(x_0,t)={e^{-t/\tau}\over \tau},
\end{equation}
where $\tau$ is the mean decay time \cite{MP},\cite{M}
($\tau(c,x_0,d)\equiv\tau_1\equiv\ae_1$):
\begin{equation}\label{tau}
\tau(c,x_0,d)=B\left\{\int\limits_{x_0}^d e^{\varphi(x)}\int\limits_{c}^x
e^{-\varphi(v)}dvdx+\int\limits_{d}^{\infty}
e^{\varphi(x)}dx \int\limits_{c}^d e^{-\varphi(v)}dv\right\}.
\end{equation}
This expression is a direct transformation of formula (\ref{t1}),
where $c$ is arbitrary, such that $c<x_0<d$.

Probably,  similar  procedure  was previously used (see \cite{H},
\cite{SA}, \cite{Roy}, \cite{LW}) for summation of
the set of moments of the First Passage Time, when exponential
distribution of the First Passage Time probability density
was demonstrated for the case of a high potential barrier
in comparison with noise intensity.

\section{Nondecay probability evolution}

Integrating probability density (\ref{11}), taking into account
definition (\ref{2a}), we get the following expression for the
nondecay probability $P(x_0,t)$ ($P(x_0,0)=1$, $P(x_0,\infty)=0$):
\begin{equation}\label{pr}
P(x_0,t)=\exp(-t/\tau),
\end{equation}
where mean decay time $\tau$ is expressed by (\ref{tau}).
Probability  (\ref{pr}) represents a well-known exponential decay
of a metastable state with a high potential barrier \cite{G}.
Where is the boundary of validity of formula (\ref{pr}) and
when can we neglect by reminders $r_n$ and $R_n$ in formulas
(\ref{9}),(\ref{10})?
To answer this question we have considered three examples
of potentials having metastable states and compared
numerically obtained nondecay probability
$P(x_0,t)=\int\limits_{c}^d W(x,t)dx$
with its exponential approximation (\ref{pr}).
We used the usual explicit difference scheme to
solve the FPE (\ref{3}), supposing the
reflecting boundary condition $G(c_b,t)=0$ ($c_b<c$) far above
the potential minimum and the absorbing one $W(d_b,t)=0$ ($d_b>d$)
far below the potential maximum, instead of boundary conditions
at $\pm\infty$, such that the influence of phantom boundaries at $c_b$
and $d_b$ on the process of diffusion was negligible.

The first considered system is described by the potential
${\mit\Phi(x)}=ax^2-bx^3$.
We have taken the following particular parameters:
$a=2$, $b=1$ that leads to the barrier height
$\Delta{\mit\Phi}\approx 1.2$, $c=-2$, $d=2a/3b$, and
$kT=0.5;1;3$. The corresponding curves of the numerically
simulated probability and its exponential approximation are
presented in Fig.1. In the worse case when $kT=1$ the maximal
difference between the corresponding curves is $3.2\%$. For
comparison, there is also presented a curve of exponential
approximation with the mean First Passage Time (MFPT) of the point
$d$ for $kT=1$ (dashed line). One can see, that in the latter
case the error is significantly larger.

The second considered system is described by the potential
${\mit\Phi(x)}=ax^4-bx^5$.
We have taken the following particular parameters:
$a=1$, $b=0.5$ that leads to the barrier height
$\Delta{\mit\Phi}\approx 1.3$, $c=-1.5$, $d=4a/5b$,
and $kT=0.5;1;3$. The corresponding curves of the numerically
simulated probability and its exponential approximation are
presented in Fig.2. In the worse case ($kT=1$) the maximal
difference between the corresponding curves is $3.4\%$.

The third considered system is described by the potential
${\mit\Phi(x)}=1-\cos(x)-ax$.
This potential is multistable. We have considered it
in the interval $[-10,10]$, taking into account three
neighboring minima.
We have taken $a=0.85$ that leads to the barrier height
$\Delta{\mit\Phi}\approx 0.1$, $c=-\pi-\arcsin(a)$,
$d=\pi-\arcsin(a)$, $x_0=\arcsin(a)$,
and $kT=0.1;0.3;1$. The corresponding curves of the
numerically simulated probability and its exponential approximation are
presented in Fig.3. In difference with two previous examples,
this potential was considered in essentially longer interval
and with smaller barrier. The difference between
curves of the numerically simulated probability and its exponential
approximation is larger. Nevertheless, the qualitative
coincidence is good enough.

Finally,  we  have  considered  an  example  of  metastable state
without potential barrier: ${\mit\Phi(x)}=-bx^3$, where $b=1$,
$x_0=-1$, $d=0$, $c=-3$ and $kT=0.1;1;5$.
By dashed curve an exponential approximation with the MFPT
of the point $d$ for $kT=1$ is presented.
It is seen, that even for such example the exponential approximation
(with the mean decay time (\ref{tau})) gives an adequate
description of the probability evolution and that this approximation
works better for larger noise intensity.

\section{Conclusion}

In the present paper the decay of metastable states,
described by the model of Markov process, has been considered.
Recurrent formulas of exact moments of decay time, valid
for arbitrary noise intensity, have been obtained.
Some concrete examples of metastable states have been analysed
numerically, and the time evolution of the nondecay probability
of a metastable state is found to be really close to the exponent
even for the case when the potential barrier height is comparable
or smaller than the noise intensity if the exact mean decay time
(\ref{tau}) is substituted into the factor of exponent.

For all investigated examples, the exponential
approximation gives an adequate behavior of the probability.
This approximation may be used in a wide range of parameters,
enough for solution of many practical tasks, but it is necessary
to remark, that the exponential approximation may lead
to a significant error in the case of extremely large noise
intensity, and in the case when the noise intensity is small,
the potential is tilted, and the barrier is absent
(purely dynamical motion slightly modulated by noise perturbations).

\section{Acknowledgments}

Authors wish to thank Prof. V.Belykh, Prof. P.Talkner, Prof. P.Jung and
Dr. W.Nadler for discussions and constructive comments.
This work has been supported by the Russian
Foundation for Basic Research (Project N~96-02-16772-a,
Project N~96-15-96718 and Project N~97-02-16928), by Ministry
of High Education of Russian Federation (Project N~3877)
and in part by Grant N~98-2-13 from the International Center
for Advanced Studies in Nizhny Novgorod.

\newpage

\begin{figure}[th]
\centerline{
\epsfxsize=14cm
\epsffile{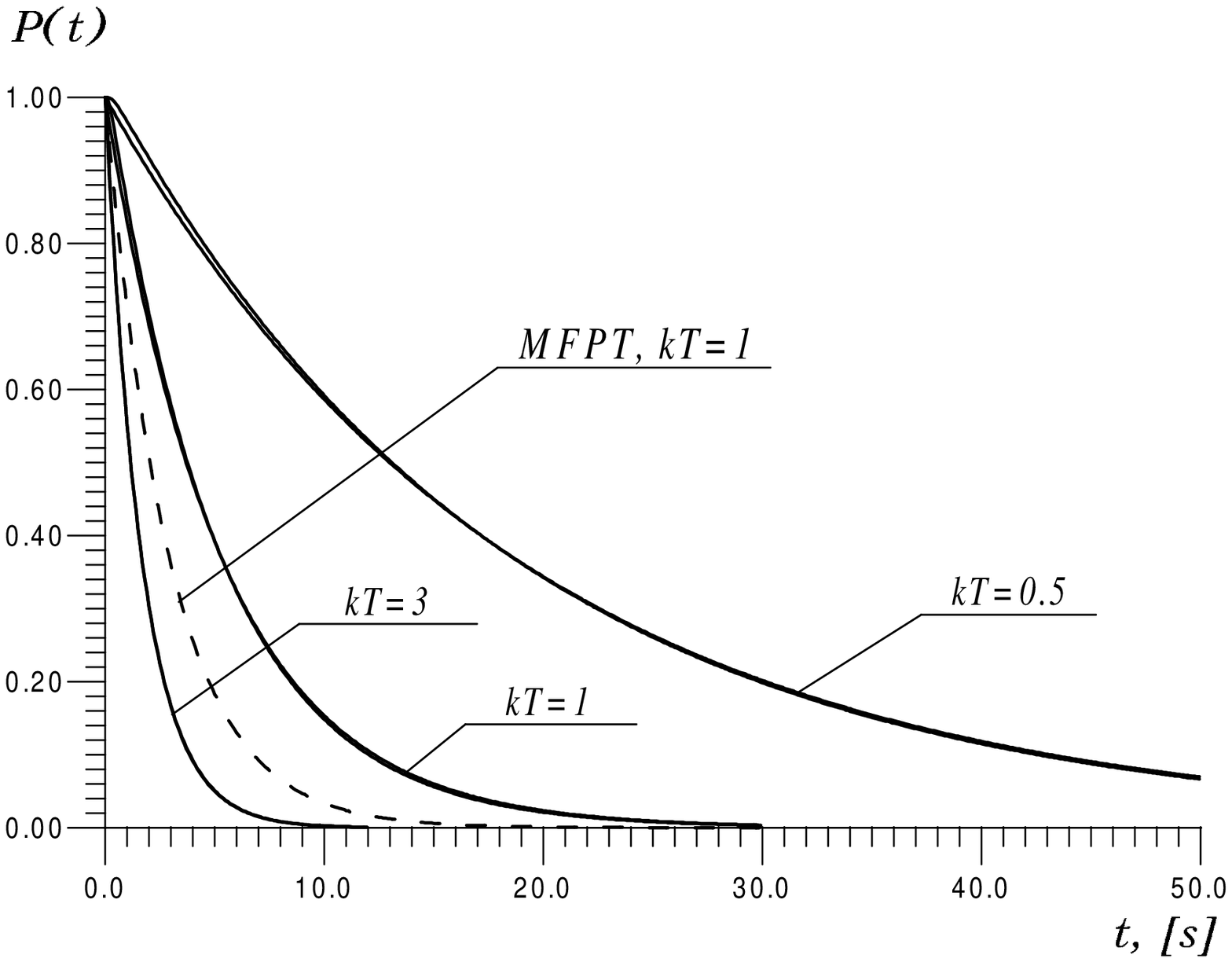}}
\caption[b]{\label{fig1}
Evolution of the nondecay probability for the potential
${\mit\Phi(x)}=ax^2-bx^3$
for different values of noise intensity; the dashed curve denoted as
MFPT (mean First Passage Time) represents exponential
approximation with MFPT substituted into the factor of exponent.}
\end{figure}

\newpage

\begin{figure}[th]
\centerline{
\epsfxsize=14cm
\epsffile{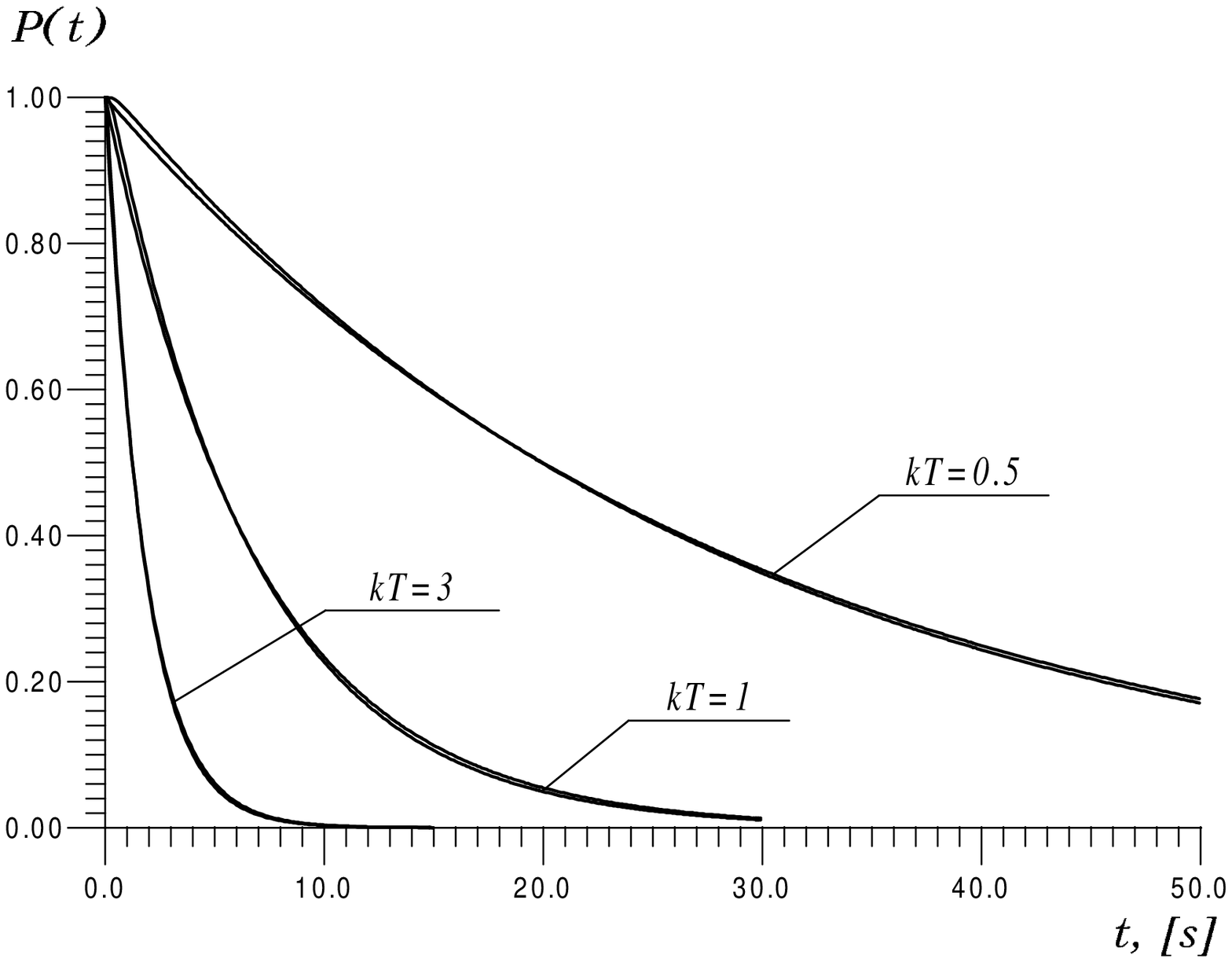}}
\caption[b]{\label{fig2}
Evolution of the nondecay probability for the potential
${\mit\Phi(x)}=ax^4-bx^5$
for different values of noise intensity.}
\end{figure}

\newpage

\begin{figure}[th]
\centerline{
\epsfxsize=14cm
\epsffile{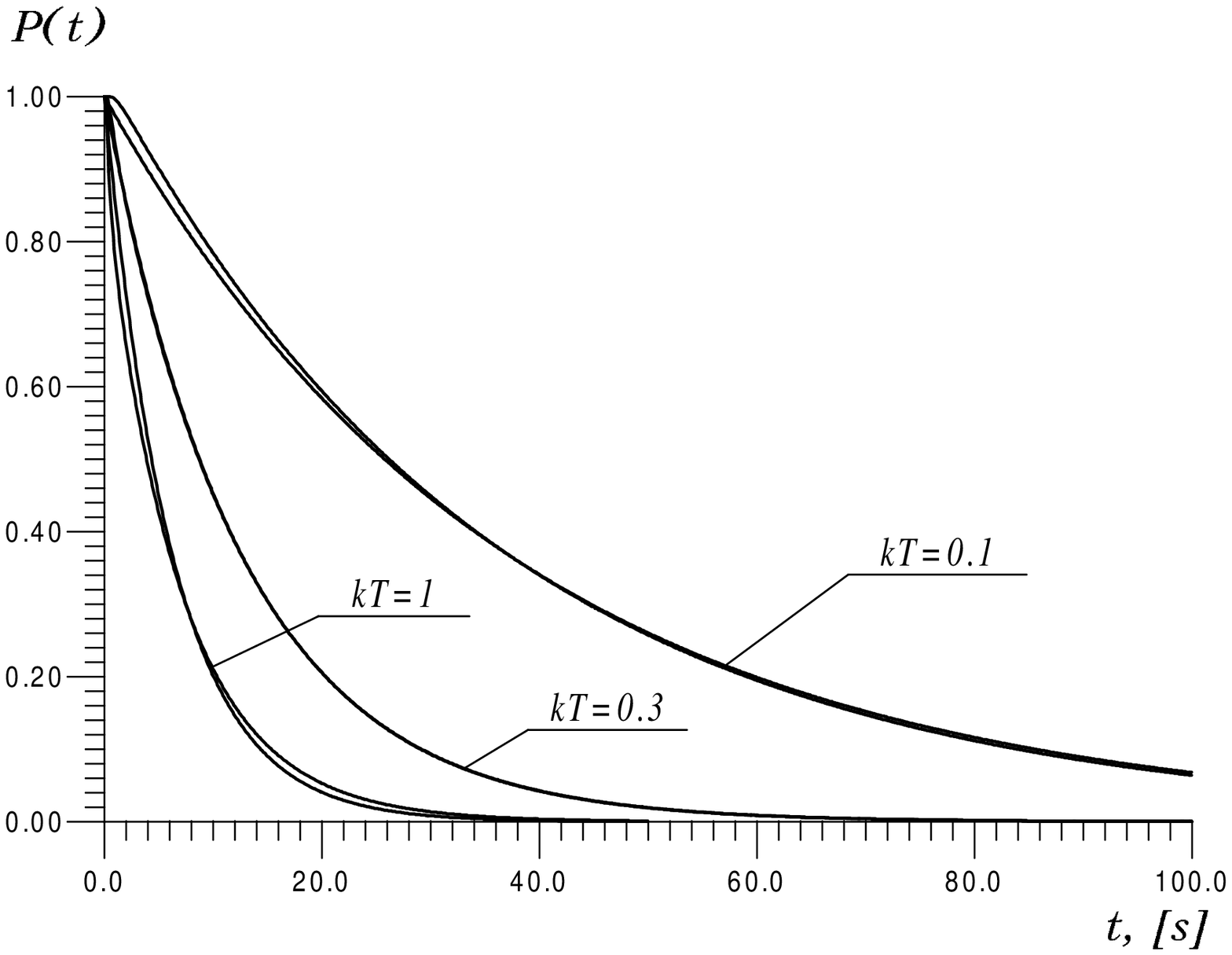}}
\caption[b]{\label{fig3}
Evolution of the nondecay probability for the potential
${\mit\Phi(x)}=1-\cos(x)-ax$
for different values of noise intensity.}
\end{figure}

\newpage

\begin{figure}[th]
\centerline{
\epsfxsize=14cm
\epsffile{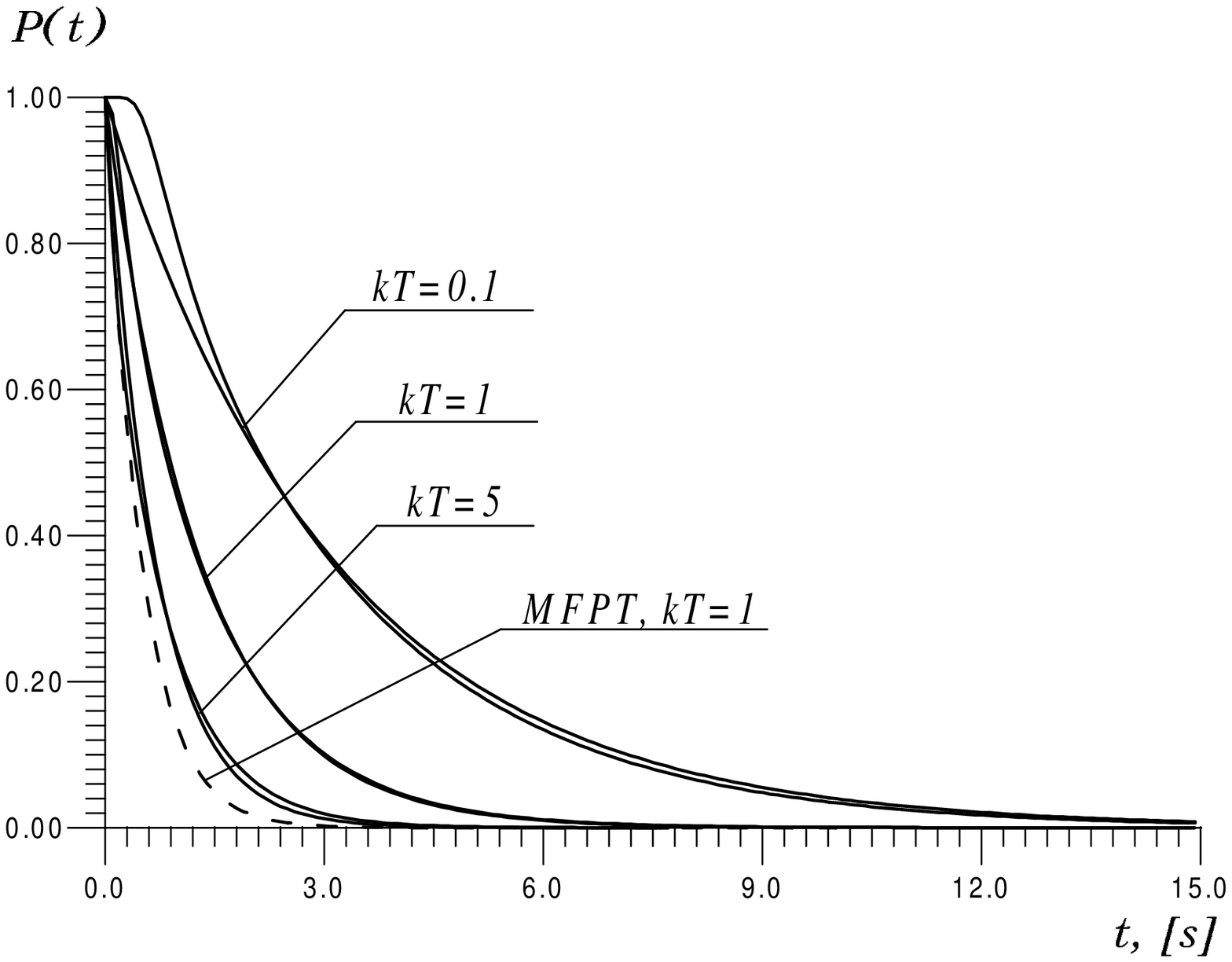}}
\caption[b]{\label{fig4}
Evolution of the nondecay probability for the potential
${\mit\Phi(x)}=-bx^3$
for different values of noise intensity; the dashed curve denoted as
MFPT (mean First Passage Time) represents exponential
approximation with MFPT substituted into the factor of exponent.}
\end{figure}

\end{document}